\documentclass[aps,pra,preprint,showpacs,superscriptaddress,groupedaddress]{revtex4}
\usepackage{amsmath}
\usepackage{tipa}
\usepackage{bbm}
\usepackage{txfonts}
\usepackage{graphicx}
\usepackage{dcolumn}
\usepackage{bm}
\usepackage{amssymb}
\usepackage{latexsym}

\begin{document}


\title{Testing the Universality of Free Fall at ${10^{ - 10}}$ level by Comparing the Atoms in Different Hyperfine States with Bragg Diffraction}
\author{Ke Zhang}
\author{Min-Kang Zhou}\email[E-mail: ]{zmk@hust.edu.cn}
\author{Yuan Cheng}
\author{Le-Le Chen}
\author{Qin Luo}
\author{Wen-Jie Xu}
\author{Lu-Shuai Cao}
\author{Xiao-Chun Duan}
\author{Zhong-Kun Hu}\email[E-mail: ]{zkhu@hust.edu.cn}

\affiliation{MOE Key Laboratory of Fundamental Physical Quantities Measurements, Hubei Key Laboratory of Gravitation and Quantum Physics, School of Physics, Huazhong University of Science and Technology, Wuhan 430074, People's Republic of China}

\date{\today}

\begin{abstract}

We have performed a precision atomic interferometry experiment on testing the universality of free fall (UFF) considering atoms' spin degree of freedom. Our experiment employs the Bragg atom interferometer with $^{87}$Rb atoms either in hyperfine state $\left| {F = 1,{m_F} = 0} \right\rangle $ or $\left| {F = 2,{m_F} = 0} \right\rangle $, and the wave packets in these two states are diffracted by one pair of Bragg beams alternatively, which can help suppress the common-mode systematic errors. We have obtained an E$\rm{\ddot{o}}$tv$\rm{\ddot{o}}$s ratio $\eta = \left( { 0.9 \pm 2.7} \right) \times {10^{ - 10}}$, and set a new record on the precision with a nearly 5 times improvement. Our experiment gives stronger restrictions on the possible UFF breaking mechanism.

\end{abstract}


\maketitle

The General Relativity (GR) has made imperial success in modern physics for describing gravity. The huge success of GR has also inspired extensive research on the extension of the theory, which bears the hope for, e.g., unifying the fundamental interactions \cite{Dam96,Cap11}. The validity of universality of free fall (UFF), being one of the fundamental postulations of GR \cite{Mis73}, has excited a huge amount of experiments \cite{Sch08,Wil04} under various circumstances to search for the sign of the extended GR theory. The most accurate tests for UFF to date were provided by the MICROSCOPE satellite mission \cite{Tou17} at the relative precision of $10^{ - 14}$ level. Other space-born experiments have also been proposed \cite{Tin13,Agu14,Bar16}.

The UFF test has also been extended to the domain of quantum technology based on matter-wave interferences \cite{Bor89,Pet99,Mer10,Sch14,Bon13,Fra04,Zho15,Tar14,Dua16,Ros17,Gei18}. Testing UFF with quantum method was performed between different atomic species like Rb and K \cite{Sch14}, or different isotopes of one species \cite{Bon13,Fra04,Zho15}. For example, a precision level of $10^{ - 8}$ was reached in an atomic fountain containing the isotope of rubidium \cite{Zho15}, and experiments with much higher precision were proposed \cite{Har15,Kov15}. Quantum test of UFF is not only advancing in the potential improvements of precision, but also particularly interesting in searching possible spin-gravity coupling and torsion of space time. Atoms possessing well defined spin properties, like the fermionic and bosonic isotopes of Sr \cite{Tar14}, the $^{87}$Rb with opposite spin orientations \cite{Dua16}, the $^{85}$Rb in different hyperfine states \cite{Fra04}, were employed as test masses in the UFF experiments. An experimental implementation using entangled atoms of $^{85}$Rb and $^{87}$Rb has also been proposed \cite{Gei18}. Especially, a relative precision of low $10^{ - 9}$ has been achieved by $^{87}$Rb atoms prepared in two hyperfine states and in their superposition \cite{Ros17}. In this letter, we present an improved UFF test at precision of $2.7\times {10^{ - 10}}$ through the comparison of the free fall of $^{87}$Rb atoms in different hyperfine states.

\begin{figure}[bbp]
\includegraphics[trim=30 10 30 30,width=0.4\textwidth]{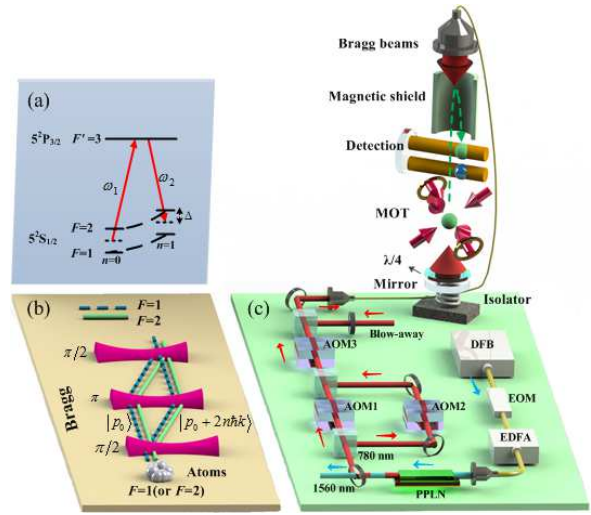}
\caption{\label{fig:1}(color online) The experimental scheme, including (a) the level scheme of the internal state-labeled Bragg diffraction, (b) the space-time diagram of the internal state-labeled Bragg atom interferometer and (c) the experimental setup with optics and atom fountain.}
\end{figure}

In this experiment, we perform Bragg interferometry measurements of the gravity acceleration difference between Rb atoms in states $\left| {5{S_{1/2}},F = 1,{m_F} = 0} \right\rangle $ and $\left| {5{S_{1/2}},F = 2,{m_F} = 0} \right\rangle $, termed as $\Delta g = {g_{F = 1}} - {g_{F = 2}}$. As shown in Fig. \ref{fig:1}(a), $^{87}$Rb atoms are initially prepared in the magnetic-insensitive state $\left| {{m_F} = 0} \right\rangle $ either populated in states $\left| {5{S_{{1 \mathord{\left/
 {\vphantom {1 2}} \right.
 \kern-\nulldelimiterspace} 2}}},F = 1} \right\rangle $ or $\left| {5{S_{{1 \mathord{\left/
 {\vphantom {1 2}} \right.
 \kern-\nulldelimiterspace} 2}}},F = 2} \right\rangle $. Provided a proper laser frequency detuning $\Delta$, both states can be coupled to the same Bragg laser beam and manipulated between the momentum states $\left| {{p_0}} \right\rangle $ and $\left| {{p_0} + 2n\hbar k} \right\rangle $, i.e. the states with momenta ${p_0}$ and ${p_0} + 2n\hbar k$ respectively, via the Bragg diffraction \cite{Mul08,Mull08,Alt13,Est15,Maz15,Ami16,Har16}. Then we can construct two Bragg atom interferometers with corresponding labeled hyperfine states in Fig. \ref{fig:1}(b), and get the interference phase ${\phi _{F = 1}}$ and ${\phi _{F = 2}}$, respectively. The free fall acceleration ${g_{F = 1}}$ (or ${g_{F = 2}}$) with atoms in different spin states is proportional to the corresponding interference phase.
 Then the UFF signal can be extracted from the phase difference (${\phi _{F = 1}} - {\phi _{F = 2}}$) of the two interferometers. The main advantages for our scheme are that, i) the Bragg atom interferometer does not change the internal hyperfine state, which makes this method intrinsically insensitive to the noise arising from the external electromagnetic field, and ii) both the two interferometers labeled by $F=1$ and $F=2$ share the same Bragg beam, therefore some systematic effects coupled with the driven laser can be significantly common rejected.

We propose that our experiment can give a new constraint on the possible breaking mechanism of UFF due to the spin degree of freedom\cite{Obu01,Pap08,Obu01,Lam06}. In our previous experiment \cite{Dua16}, we have performed tests on the UFF between $^{87}$Rb in states of $\left| {5{S_{1/2}},{m_F} = + 1} \right\rangle $ and $\left| {5{S_{1/2}},{m_F} = - 1} \right\rangle $, which corresponds to the test of the possible UFF breaking mechanism due to the spin projected to the direction of the gravity force. There still remains an open question, whether the spin projected to the perpendicular plane of the gravity force can break the UFF or not. Our current experiment addresses this open question and contributes to a more complete knowledge on the breaking of UFF due to the spin degree of freedom. To test the possible breaking of UFF against the spin projected to the plane perpendicular to the gravity force, it is natural to assume that the breaking effect only depends on the amplitude but not the polarization of the spin in the plane, $e.g.$ assuming a rotation symmetry in the perpendicular plane. A Hamiltonian term corresponding to such a breaking mechanism reads:
\begin{equation}
{V_ \bot }(z) = \tilde k{\left| {{F_ \bot }} \right|^2}mgz,\label{Eq:1}
\end{equation}
where the gravity force is taken along the z direction, and ${\left| {{F_ \bot }} \right|^2}$ is the amplitude of the spin projection in the horizontal plane of the gravity force which equals to 2 and 6 for $F=1$ and $F=2$. Our experiment can then give an upper bound of $\left( {-0.2 \pm 0.7} \right) \times {10^{ - 10}}$ to the strength of the UFF breaking term $\tilde k$. Because atoms are prepared in different hyperfine states with different internal energy, our experiment also corresponds to a test of the diagonal terms of the possible breaking operator of the UFF as ${r_1} - {r_2}{\rm{ = }}(0.9 \pm 2.7) \times {10^{{\rm{ - 10}}}}$ according to the mass-energy equivalence, which correspond to an improvement over previous results of about a factor of 5\cite{Ros17,Zyc17}.

Firstly, we offer a detailed description of the experiment setup. The key point for realizing two Bragg atom interferometers with different hyperfine states is that, we have to ensure the effective Rabi frequency ${\Omega _{{\rm{eff}}}}$ of $F=1$ equals to that of $F=2$ when both of them couple to the same Bragg laser beams. For the Gaussian shape Bragg pulses, the nth-order effective Rabi frequency \cite{Mul08} also depends on the normal two-photon Rabi frequency $\Omega $, where $\Omega $ is inversely proportional to the single photon frequency detuning $\Delta $ \cite{Kas92}. We denote this two photon Rabi frequency as ${\Omega _1}$ and ${\Omega _2}$ for $F=1$ and $F=2$ state, respectively. For the requirements of ${\Omega _1} = {\Omega _2}$, and considering the couplings of hyperfine states in $\left| {5{P_{{3 \mathord{\left/
 {\vphantom {3 2}} \right.
 \kern-\nulldelimiterspace} 2}}}} \right\rangle $, the detuning $\Delta $ should be set around 3.1817 GHz. In this configuration, Bragg beams are red (blue) detuned for $F=1$ ($F=2$) state.

Atom interferometer with Bragg diffraction requires high power laser beams. The frequency doubling method is employed to produce more than 1 W Bragg laser beam at 780 nm (Fig. \ref{fig:1}(c)). A narrow-linewidth distributed feedback (DFB) seed laser at the telecom wavelength is amplified to 30 W by an erbium-doped fiber amplifier (EDFA, IPG photonics). Then the output beam from the EDFA passes through a periodically poled lithium niobate (PPLN) crystal, which can double the laser frequency from 1560 nm to 780 nm \cite{San12}. The 780 nm laser is split into two beams (beam1 and beam2) and are frequency shifted by two pieces of acousto-optical modulators (AOM1 and AOM2) respectively. The two counter-propagating Bragg beams are composed by beam 1 and 2 with perpendicular polarization. The frequency difference of the two beams is adjusted by either AOM1 or AOM2 to match the resonance condition, which is noted as $\Delta \omega = {\omega _1} - {\omega _2} = 2{\bf{k}} \cdot {{\bf{v}}_a} + 4n{\omega _r}$, where ${\omega _r}$ is the single photon recoil frequency shift. The Doppler frequency shift $2{\bf{k}} \cdot {{\bf{v}}_a}$ due to free fall can also be compensated by one of the AOM, where ${\bf{k}}$ and ${{\bf{v}}_a}$ represent the wave vector of single Bragg beam and the atom's free fall velocity, respectively. In order to maximize the diffraction efficiency, the shape of Bragg pulses is programmed to a Gaussian form \cite{Mull08} according to AOM3. Before injecting into the vacuum chamber, about 80 mW Bragg beams are overlapped with the blow-away beams in a single mode polarization-maintaining (PM) fiber. The Raman beams which are employed on the velocity selection in the vertical direction are produced by a fiber electro-optic modulator (EOM) with a source of 6.83 GHz before the EDFA. With shutting down the AOM2, they pass through the AOM1 and AOM3. So the Bragg beams and Raman beams share a same optical path and are able to switch alternately by turning on or off the driven sources of EOM and AOM2. The ${e^{ - 2}}$ diameter of both the Bragg and Raman beams is about 19 mm. All of the beams are aligned and injected into the vacuum chamber through the top window, passing through a quarter wave plate, and retro-reflected by a reference mirror on a vibration isolator.

The interference for matter-wave is implemented based on a cold $^{87}$Rb atom fountain that can be found elsewhere \cite{Hu13}. The total height of the fountain is 0.66 m. The initial state preparation that promises atoms either in $\left| {F = 1,{m_F} = 0} \right\rangle $ or in $\left| {F = 2,{m_F} = 0} \right\rangle $ is necessary before they fly into the interferometer chamber. This is realized with the microwave $\pi$-pulses between the hyperfine states, the repumping laser, the blow away beam of lower state ($\left| {5{S_{{1 \mathord{\left/
 {\vphantom {1 2}} \right.
 \kern-\nulldelimiterspace} 2}}},F = 1} \right\rangle $ to $\left| {5{P_{{3 \mathord{\left/
 {\vphantom {3 2}} \right.
 \kern-\nulldelimiterspace} 2}}},F = 0} \right\rangle $), and the blow away beam of upper state ($\left| {5{S_{{1 \mathord{\left/
 {\vphantom {1 2}} \right.
 \kern-\nulldelimiterspace} 2}}},F = 2} \right\rangle $ to $\left| {5{P_{{3 \mathord{\left/
 {\vphantom {3 2}} \right.
 \kern-\nulldelimiterspace} 2}}},F = 3} \right\rangle $). By compositing the above microwave pulses and the blow away laser beams, we can prepare atoms into the two target states alternately shot by shot. A Doppler-sensitive Raman $\pi$-pulse 80 $\mu$s long further prepares atoms in a narrow vertical momentum width less than 0.37 $\hbar k$.

\begin{figure}[bbp]
\includegraphics[trim=40 10 40 20,width=0.40\textwidth]{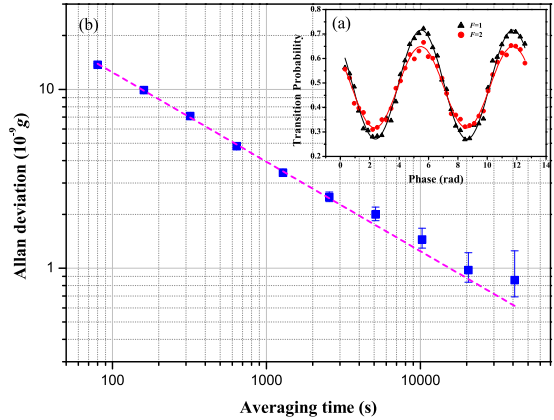}
\caption{\label{fig:2}(color online) (a) Fringes of the two Bragg atom interferometers. The black triangles (red dots) are experimental data for atoms in $F=1$ ($F=2$), and the black line (red line) represents a sine curve fitting. (b) Allan deviation of the differential acceleration measurements. The short-term sensitivity is $1.2 \times {10^{ - 7}}g$/Hz${}^{1/2}$ from the $t^{-1/2}$ fitting (magenta line).}
\end{figure}

 After atoms enter in to the magnetic shielding region, a series of Bragg pulse in the form of ${\pi \mathord{\left/
 {\vphantom {\pi 2}} \right.
 \kern-\nulldelimiterspace} 2} - \pi - {\pi \mathord{\left/
 {\vphantom {\pi 2}} \right.
 \kern-\nulldelimiterspace} 2}$ manipulate the atomic wavepacket regardless of their internal states. All the Bragg pulses are programmed to a Gaussian shape. The typical width for the $\pi $ pulse is about 42 $\mu$s. A diffraction order $n=1$ is selected for both states by carefully setting $\Delta \omega = 2{\bf{k}} \cdot {{\bf{v}}_a} + 4{\omega _r}$. The laser frequency detuning $\Delta $ is well adjusted, thus interferences can happen between $\left| {F = 1,{p_0}} \right\rangle $($\left| {F = 2,{p_0}} \right\rangle $) and $\left| {F = 1,{p_0} + 2\hbar k} \right\rangle $($\left| {F = 2,{p_0} + 2\hbar k} \right\rangle $). The first order diffraction efficiency both for $\left| {F = 1} \right\rangle $ and $\left| {F = 2} \right\rangle $ can reach $88\% $ with a single Bragg $\pi - $pulse. The free evolution time $T$ between two Bragg pulses is 150 ms.

When atoms fall back to the detection region, the two atomic clouds
in the two interference paths are still partly overlapped in
vertical direction, and can't be easily distinguished with the
normal time of flight method. Here we use the Doppler-sensitive
Raman spectroscopy method \cite{Kas91,chengyuan18} with Raman $\pi$
pulses to get the population of $\left| {{p_0}} \right\rangle $ and
$\left| {{p_0} + 2\hbar k} \right\rangle $ in momentum space through
a fluorescence measurement. The frequency difference between these
two states due to the Doppler effect in the spectrum is about 30
kHz, while the resolution of our Raman spectroscopy can be better
than 0.3 kHz, which is good enough to distinguish the $\left|
{{p_0}} \right\rangle $ and $\left| {{p_0} + 2\hbar k} \right\rangle
$ states and measure their population. For the Bragg atom
interferometer labeled by $\left| {F = 1} \right\rangle $, the Raman
spectrum is obtained by sweeping the Raman laser's frequency, and
the probability of finding atoms in state $\left| {F = 1,{p_0} +
2\hbar k} \right\rangle $ is given by the amplitude ratio of the two
peaks corresponding to $\left| {F = 1,{p_0}} \right\rangle $ and
$\left| {F = 1,{p_0} + 2\hbar k} \right\rangle $. The two peaks can
be found with two fixed Raman frequencies and the probability is
primarily sensitive to the peak values, therefore two measurements
with two shots are required to get one probability. Each measurement
including the MOT loading, state preparation, interference stage and
detection takes 1 s. The detection for the momentum states labeled
by $\left| {F = 2} \right\rangle $ is also performed with the same
strategy.

The probability of finding atoms in $\left| {{p_0} + 2\hbar k} \right\rangle $ state depends on the interferometry phase, and can be written as $P = {{\left( {1 - \cos \left( {n\left( {{k_{{\rm{eff}}}}g - \alpha } \right){T^2}} \right)} \right)} \mathord{\left/
 {\vphantom {{\left( {1 - \cos \left( {n\left( {{k_{{\rm{eff}}}}g - \alpha } \right){T^2}} \right)} \right)} 2}} \right.
 \kern-\nulldelimiterspace} 2}$. Here $\alpha $ is Bragg beam's frequency chirp rate for compensating the Doppler shift, the effective wave vector ${\vec k_{{\rm{eff}}}} = {\vec k_1} - {\vec k_2}$ relies on the wave number of upper and down shooting Bragg beams. The matter-wave interference is performed either in state $F=1$ or $F=2$ by using Bragg diffraction with a time separation of 2 s alternately. As shown in Fig. \ref{fig:2}(a), fringes for $F=1$ and $F=2$ with similar contrast are obtained by slightly modulating the driven frequency of AOM2. Each fringe contains two periods corresponding to a phase interval of $4\pi $ and taking 160 s totally.

The differential acceleration ($\Delta g={g_{F = 1}} - {g_{F = 2}}$) in Fig. \ref{fig:2}(a) is used to test the UFF with atoms in different hyperfine states. Fig. \ref{fig:2}(b) shows the Allan deviation of the differential acceleration measurements by this state-labeled Bragg atom interferometer. The short-term sensitivity of $\Delta g$ is $1.2 \times {10^{ - 7}}g/{\rm{H}}{{\rm{z}}^{{1 \mathord{\left/
 {\vphantom {1 2}} \right.
 \kern-\nulldelimiterspace} 2}}}$. The resolution of the differential measurement scales at $t^{-1/2}$, and can be better than $1 \times {10^{ - 9}}g$ at 20000 s, which promise a UFF test with our scheme at $10^{-10}$ level .

\begin{figure}[tbp]
\includegraphics[trim=40 10 40 20,width=0.40\textwidth]{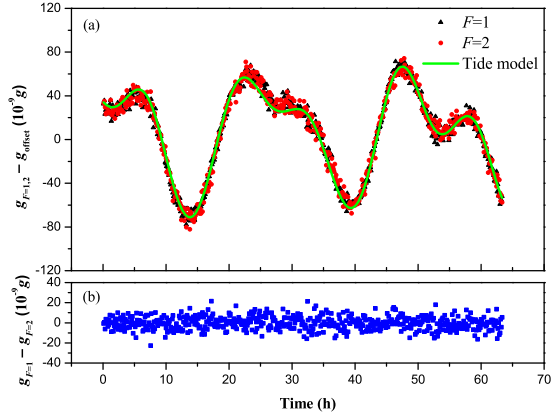}
\caption{\label{fig:3}(color online) The UFF test within 63 hours measurements. (a) The gravity acceleration measured with atoms in $F=1$ (black triangles) and $F=2$ (red dots) comparing to the tide model (green line). (b) The differential acceleration of the two states (blue squares).}
\end{figure}

A test of the UFF is then performed by continuously measuring the gravity acceleration with this state-labeled atom interferometer. As shown in Fig. \ref{fig:3}, about 63 hours data is recorded by the apparatus. Each point in this data is the mean result of 400 s. Both the two interferometers with $\left| {F = 1} \right\rangle $ and $\left| {F = 2} \right\rangle $ can precisely map the gravity tides, which are displayed in Fig. \ref{fig:3}(a) by subtracting a constant offset ${g_{{\rm{offset}}}}$. What we care about in the UFF test is the differential acceleration $\Delta g$ shown by blue squares in Fig. \ref{fig:3}(b). By averaging all the data, we get $\Delta g = \left( { - 1.2 \pm 2.6} \right) \times {10^{ - 10}}g$ where the uncertainty is the standard deviation of the weighted mean.

In our experiment, atoms in $F=1$ and $F=2$ are prepared in the same way and only one pair of Bragg beams is employed to diffract atoms both in $F=1$ and $F=2$. Therefore some systematic effects can be common rejected by differential measurement, such as the gravity gradient effect, the Coriolis effect and the wavefront aberration. The fluctuations of these effects between two states due to the alternatively measurements only contribute to the noise of the differential measurement which is included in the statistical uncertainty. The main systematic effects for $\Delta g$ are listed in Table \ref{tab:table1}. The magnetic field inhomogeneity contribute with a bias to $\Delta g$, because of the spatial separation of the two arms of the Bragg atom interferometer and the opposite sign of the Land\'{e} g-factor for $F=1$ and $F=2$ state. As atoms are prepared in ${m_F} = 0$ state, we only have to consider the quadratic Zeeman effect. The magnetic field in the interferometry region is measured precisely by the Raman spectroscopy method \cite{Zho10}, and the Zeeman effect on the UFF test due to spatial separation of wavepacket is evaluated to be $\left( { -2.1 \pm 0.5} \right) \times {10^{ - 10}}g$ giving a dominant systematic impact. This effect is also confirmed with modulation experiments by measuring the differential acceleration $\Delta g$ in different magnetic bias field. As shown in Fig.4, when increasing the magnetic field, the value of $\Delta g$ performs as a quadratic increase, which is consistent with the evaluation values based on the magnetic field distribution. Because atoms are in a same internal state for Bragg type interferometer, the ac Stark shifts caused by the spatial intensity gradients of the Bragg lasers \cite{Ros17}and the intensity fluctuation between the first and third Bragg pulses \cite{Alt13} are both less than $1 \times {10^{ - 11}}g$ in our experiment. Limited by the accuracy of absolute frequency of the Bragg lasers, the maximum frequency deviation of 1 MHz from the detuning $\Delta=3.1817$ GHz will contribute less than $1 \times {10^{ - 11}}g$ due to the two-photon light shift \cite{Gau08,Gie16}. The local gravity variation due to the tides will induce a systematic error as the measurement in $F=1$ is always 2 s after $F=2$. This effect is evaluated at the level of $3 \times {10^{ - 12}}g$ and can be neglected in the present test. As we select the first order Bragg diffraction to manipulate atoms without other unwanted momentum states, there should be no parasitic interference \cite{Est15,Par16}.

\begin{figure}[tbp]
\includegraphics[trim=40 10 40 20,width=0.40\textwidth]{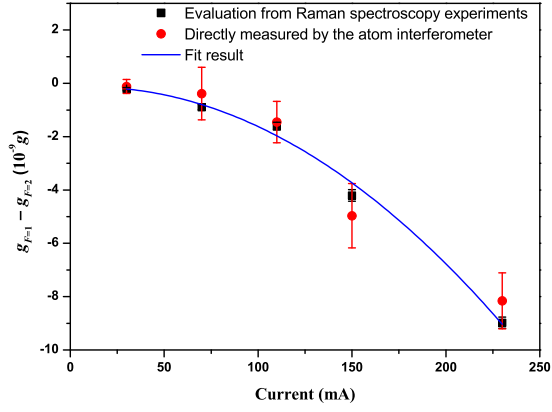}
\caption{\label{fig:4}(color online) Measurements of the quadratic Zeeman effect by modulating the current in the bias solenoid. In different currents, the values directly measured by the atom interferometers (red dots) agree with the evaluation values based on the magnetic field distribution from Raman spectroscopy experiments (black squares). The blue line is the quadratic polynomial fit result.Here in our apparatus, the bias magnetic field is 90 mG when the current is 100 mA.}
\end{figure}

After considering and correction of these systematic effects described above which are summarized in Table \ref{tab:table1} , the E$\rm{\ddot{o}}$tv$\rm{\ddot{o}}$s ratio given by
\begin{equation}
{\eta _{1 - 2}} = 2{{\left( {{g_{F = 1}} - {g_{F = 2}}} \right)} \mathord{\left/
 {\vphantom {{\left( {{g_{F = 1}} - {g_{F = 2}}} \right)} {\left( {{g_{F = 1}} + {g_{F = 2}}} \right)}}} \right.
 \kern-\nulldelimiterspace} {\left( {{g_{F = 1}} + {g_{F = 2}}} \right)}}\label{Eq:2}
\end{equation}
is finally determined to be ${\eta _{1 - 2}} = \left( {0.9 \pm 2.7} \right) \times {10^{ - 10}}$, which means the UFF between atoms in different hyperfine states is still valid at the precision of $10{^{-10}}$ level. Quantitatively, a direct upper bound of $\tilde k$ is given by $\tilde k = - ({g_{F = 2}} - {g_{F = 1}})/4g = - \eta /4 = \left( { - 0.2 \pm {\rm{0}}{\rm{.7}}} \right) \times {10^{{\rm{ - 10}}}}$. The diagonal terms of the possible breaking operator of a UFF violation is also estimated to be ${r_1} - {r_2}{\rm{ = }}(0.9 \pm 2.7) \times {10^{{\rm{ - 10}}}}$.

\begin{table}
\caption{\label{tab:table1}Main contributions to the differential acceleration measurements. }
\begin{ruledtabular}
\begin{tabular}{lcr}
&$\Delta g$($\times {10^{ - 10}}g$) & Uncertainty($ \times {10^{ - 10}}g$)\\
\hline
Statistical uncertainty & -1.2 & 2.6\\
\hline
Quadratic Zeeman shift & -2.1 & 0.5\\
AC Stark shift & 0 & $<$0.2\\
Tide effect & 0 & 0.03\\
Corrected & 0.9 & 2.7\\
\end{tabular}
\end{ruledtabular}
\end{table}

In conclusion, we have realized the Bragg atom interferometers with different hyperfine states, and demonstrated its application in high precision measurements of gravitational acceleration. Due to the property of coupling to the same Bragg beams, various systematic effects can be common rejected in the present precision. With this state-labeled $^{87}$Rb atom interferometer, a precise quantum test on the UFF between different hyperfine states is performed at ${10^{ - 10}}$ level, gain about 5 times improvements on the accuracy and still see no violation of UFF. The experimental scheme demonstrated here can be further developed to construct two atom interferometers simultaneously, which can be applied in the UFF test with the isotope of rubidium or other species \cite{Tar14,Zho15}, paving a way for high precision quantum test of UFF better than ${10^{ - 10}}$ level.

The authors gratefully acknowledge \v{C}aslav Brukner for the inspiring discussions on this work. This work is supported by the National Natural Science Foundation of China (Grants Nos. 11625417, 91636219, 11727809, 91736311 and 11474115).


\end{document}